\documentstyle[a4p,12pt,oldlfont,twoside,cite,epsfig,amsfonts,amsthm,amsbsy]{article}
\newcommand{\aem    }{\mbox{$\alpha$}}
\newcommand{\aemsq  }{\mbox{$\aem^2$}}
\newcommand{\invpb  }{\mbox{\rm pb$^{-1}$}}
\newcommand{\Z      }[4]{\mbox{$#1\,\pm #2\,^{+\,#3}_{-\,#4}$}}

\newcommand{\qnq    }{\mbox{$Q^{2}_{0}$}}
\newcommand{\qsq    }{\mbox{$Q^{2}$}}
\newcommand{\qsqmin }{\mbox{$Q_{\rm min}^2$}}
\newcommand{\qsqmax }{\mbox{$Q_{\rm max}^2$}}

\newcommand{\psq    }{\mbox{$P^{2}$}}

\newcommand{\ft     }{\mbox{$F_{2}^{\gamma}$}}

\newcommand{\ftn    }{\mbox{$F_{2}^{\gamma}/\aem$}}
\newcommand{\fl     }{\mbox{$F_{\rm L}^{\gamma}$}}
\newcommand{\ftxq   }{\mbox{$\ft(x,\qsq)$}}

\newcommand{\flxq   }{\mbox{$\fl(x,\qsq)$}}
\newcommand{\ftx    }{\mbox{$\ft(x)$}}

\newcommand{\gev    }{\mbox{$\rm GeV$}}
\newcommand{\gevsq  }{\mbox{$\rm GeV^2$}}
\newcommand{\zn     }{\mbox{$\rm Z^0$}}
\newcommand{\epm    }{\mbox{$\rm e^\pm$}}
\newcommand{\epem   }{\mbox{$\rm e^+e^-$}}

\newcommand{\znhad  }{\mbox{$\zn\rightarrow$ hadrons}}

\newcommand{\th     }{\mbox{$\theta$}}
\newcommand{\ph     }{\mbox{$\phi$}}
\newcommand{\etag   }{\mbox{$E_{\rm tag}$}}
\newcommand{\ea     }{\mbox{$E_{\rm at}$}}

\newcommand{\pbal   }{\mbox{$p_{\rm t}^{\rm bal}$}}
\newcommand{\pout   }{\mbox{$p_{\rm t}^{\rm out}$}}

\newcommand{\eb     }{\mbox{$E_{\rm b}$}}
\newcommand{\ttag   }{\mbox{$\theta_{\rm tag}$}}
\newcommand{\qzm    }{\mbox{$\langle \qsq \rangle$}}
\newcommand{\pzm    }{\mbox{$\langle \psq \rangle$}}

\newcommand{\Wvis   }{\mbox{$W_{\rm vis}$}}

\newcommand{\xvis   }{\mbox{$x_{\rm vis}$}}
\def\etal{{\it et al.}}
\begin{document}
\begin{titlepage}
\begin{center}{\large   EUROPEAN LABORATORY FOR PARTICLE PHYSICS
}\end{center}\bigskip
\begin{flushright}
       CERN-PPE/97-103   \\ 30 July 1997
\end{flushright}
\bigskip\bigskip\bigskip\bigskip\bigskip
\begin{center}{\LARGE\bf 
\boldmath
 Measurement of the\vspace*{0.4 cm}\linebreak
 Photon Structure Function 
 \ft\ at Low $x$
}\end{center}\bigskip\bigskip
\begin{center}{\LARGE The OPAL Collaboration
}\end{center}\bigskip\bigskip
\bigskip\begin{center}{\large  Abstract}\end{center}
 Deep inelastic electron-photon scattering is studied using \epem~data collected by the
 OPAL detector at centre-of-mass energies $\sqrt{s_{\mbox{\tiny ee}}}\approx M_{\mbox{\tiny Z$^0$}}$.
 The photon structure function \ftxq\ is explored in a \qsq~range
 of 1.1~to~6.6~\gevsq\ at lower $x$~values than ever before. 
 To probe this kinematic region events are selected with a beam electron scattered
 into one of the OPAL luminosity calorimeters 
 at scattering angles between 27 and 55~mrad. 
 A measurement is presented of the photon structure
 function \ftxq\ at \qzm~=~1.86~\gevsq\ and 3.76~\gevsq\ in five logarithmic $x$ bins from 0.0025~to~0.2. 
\bigskip\bigskip\bigskip\bigskip
\bigskip\bigskip
\begin{center}{\large
(Submitted to Physics Letters B)
}\end{center}
\end{titlepage}
\begin{center}{\Large        The OPAL Collaboration
}\end{center}\bigskip
\begin{center}{
K.\thinspace Ackerstaff$^{  8}$,
G.\thinspace Alexander$^{ 23}$,
J.\thinspace Allison$^{ 16}$,
N.\thinspace Altekamp$^{  5}$,
K.J.\thinspace Anderson$^{  9}$,
S.\thinspace Anderson$^{ 12}$,
S.\thinspace Arcelli$^{  2}$,
S.\thinspace Asai$^{ 24}$,
D.\thinspace Axen$^{ 29}$,
G.\thinspace Azuelos$^{ 18,  a}$
% ...
}\end{center}\bigskip
\bigskip
% ...
$^{  2}$Dipartimento di Fisica dell' Universit\`a di Bologna and INFN,
I-40126 Bologna, Italy
\newline
% ...
$^{  5}$Cavendish Laboratory, Cambridge CB3 0HE, UK
\newline
% ...
$^{  8}$CERN, European Organisation for Particle Physics,
CH-1211 Geneva 23, Switzerland
\newline
$^{  9}$Enrico Fermi Institute and Department of Physics,
University of Chicago, Chicago IL 60637, USA
\newline
% ...
$^{ 12}$Indiana University, Department of Physics,
Swain Hall West 117, Bloomington IN 47405, USA
\newline
% ...
$^{ 16}$Department of Physics, Schuster Laboratory, The University,
Manchester M13 9PL, UK
\newline
% ...
$^{ 18}$Laboratoire de Physique Nucl\'eaire, Universit\'e de Montr\'eal,
Montr\'eal, Quebec H3C 3J7, Canada
\newline
% ...
$^{ 23}$Department of Physics and Astronomy, Tel Aviv University,
Tel Aviv 69978, Israel
\newline
$^{ 24}$International Centre for Elementary Particle Physics and
Department of Physics, University of Tokyo, Tokyo 113, and
Kobe University, Kobe 657, Japan
\newline
% ...
$^{ 29}$University of British Columbia, Department of Physics,
Vancouver BC V6T 1Z1, Canada
\newline
% ...
$^{  a}$ and at TRIUMF, Vancouver, Canada V6T 2A3
\newline
% ...
\newline\bigskip\bigskip

%
%
%
%
%
%
 The hadronic structure function \ft\ of the photon has been measured in
 deep inelastic electron-photon scattering at various \epem~experiments (\cite{ref:PLUTO,ref:TPC,ref:TASSO,ref:f2opal} and references therein). In the previous OPAL analysis~\cite{ref:f2opal}  
 \ftxq\ was measured using tagged electrons with a range of scattering angles from 60~mrad up to 500~mrad to the initial beam direction.
 This letter describes an analysis of data collected in 1993 and 1994 at \epem\ center-of-mass energies between 89.2~GeV and 93.2~GeV, 
 using the OPAL silicon tungsten (SW) luminometer as
 the electron tagger with a clear angular acceptance from 27~mrad to 55~mrad. This corresponds to 
 a lowest \qsq\ value of 1.14~\gevsq\ in the selected sample, 
 close to the lower limit at which perturbative QCD can be expected to work.
 The energy of the LEP beams is higher than at any previous \epem\ collider so lower values of the scaling variable $x$
 can be reached at any given \qsq.
 The results reported here go down to $x = 0.0025$. This kinematical
 region is of particular interest because theoretical
 predictions differ significantly from each other~\cite{ref:GRV,ref:DG,ref:SaS1D,ref:GS,ref:WHIT,ref:LAC,ref:FKP}
 and because in this region the proton structure function is observed to start to increase as $x$ decreases~\cite{ref:HERA}.\par
 In the singly-tagged regime, with one scattered electron tagged in the detector, the two photon process 
 can be regarded as the 
 deep inelastic scattering of an \epm\ with four-momentum $k$ on a quasi-real 
 photon with four-momentum $p$.
 The flux of quasi-real photons can be calculated using the
 Equivalent Photon Approximation~\cite{ref:Weiz}.
 The cross-section for deep inelastic electron-photon scattering with a hadronic final state X
 is expressed as~\cite{ref:Berger},
 \begin{equation}
 \frac{{\rm d}^2\sigma_{\rm e\gamma\rightarrow \rm e X}}{{\rm d}x\thinspace{\rm d}Q^2}
 =\frac{2\pi\aemsq}{x\,Q^{4}}
 \left[\left( 1+(1-y)^2\right) \ftxq - y^{2} \flxq\right]\, ,
 \label{eqn:Xsect}
 \end{equation}
 where $\qsq\equiv-q^2$ is the negative four-momentum squared of the 
 virtual photon and  \aem\ is the fine structure constant.
 The usual dimensionless variables of deep inelastic 
 scattering, $x$ and $y$, are defined as 
 \begin{equation}
 x\equiv\frac{\qsq}{2\ p\cdot q}\, , \,\,\,\,\,\,\, y\equiv\frac{p\cdot q}{p\cdot k}.
 \label{eqn:xy}
 \end{equation}
 In the kinematic regime studied here ($y^2\ll 1$) the contribution 
 of the term proportional to the longitudinal structure function \flxq\ is small and is therefore neglected.
 The structure function formalism of deep inelastic scattering implies that the virtual photon behaves as a pointlike 
 probe. However, at very low \qsq\ it can also show a hadronic structure.
 For this analysis, such a hadronic contribution is neglected.\par
 The scattered electron is detected with the silicon tungsten calorimeters (SW)\cite{ref:SW}
 that are placed around the beam pipe at a distance of 2.4~m in $z$ from the interaction point on both sides of the OPAL detector, covering polar angles $\theta$ from 25~to~59~mrad\footnote{A right-handed coordinate system is used. The 
       $x$-axis points towards the 
       centre of the LEP ring, the $y$-axis upwards and the $z$-axis in
       the direction of the electron beam.  The polar angle \th\ and the 
       azimuthal angle \ph\ are defined with respect to the $z$-axis
       and $x$-axis, respectively.}.
 They are cylindrical sampling calorimeters
 consisting of 19 layers of silicon detectors interleaved with 18 layers of tungsten,
 equivalent to a total of 22 radiation lengths.
 For electromagnetic showers an energy resolution of $24\%/\sqrt{E\, [\gev]}$  and a resolution in polar angle of
 0.06~mrad is achieved.
 The hadronic final state X is measured with SW, the forward detectors FD, the OPAL electromagnetic calorimeter and
 the OPAL tracking system which consists of a silicon microvertex detector, a drift
 chamber vertex detector, a jet chamber and z-chambers\cite{ref:opalnim}.\par 
 The measurement of \ftxq\ involves the determination of \qsq\ and $x$
 that can be obtained from the four-vectors of the tagged
 electron and the hadronic final state:
%
 \begin{equation}
 \qsq\, \approx 2\,\eb\,\etag\,(1- \cos\ttag) \,\,\,\,\,\,\,\,\,\,\,\, (\mbox{\rm neglecting the electron mass})
 \end{equation}
 \begin{equation}
 x \approx \frac{\qsq}{\qsq+W^2} \,\,\,\,\,\,\,\,\,\,\,\, \mbox{\rm (for $\psq\equiv -p^2 \approx 0$)}. 
 \label{eqn:Xcalc}
 \end{equation}
%
 \etag\ and \ttag\ are the energy and polar angle of the observed 
 electron, \eb\ is the beam energy, and $W$ the invariant mass of the hadronic 
 final state.
 In addition to the tag requirement, an antitag condition is applied to ensure that the virtuality of the quasi-real photon
\psq\ is small
 and requirements are imposed on the hadronic final state to reject residual background.
%
%
%
 \begin{enumerate}
 \item A tagged electron is required, identified as a 
       cluster in the silicon tungsten calorimeter with energy 
       $0.775\,\eb\le\etag\le 1.2\,\eb$ and a polar angle 
       $27 \le \ttag \le 55$~mrad with respect to
       the beam axis, defining a lower limit of \qsqmin $=1.14$ \gevsq\ and
       an upper limit of \qsqmax $=6.57$ \gevsq\\
       (tag requirement).
 \item The energy \ea\ of the most energetic cluster in the hemisphere 
       opposite to the one containing the tagged electron is restricted to
       $\ea\le 0.25\,\eb$\\
       (antitag requirement).
 \item The visible invariant mass \Wvis\ of the hadronic system is required  to be in 
       the range
       $2.5\,\gev\le\Wvis\le 40\,\gev$.
       \Wvis\ is defined by all tracks of charged particles and all calorimeter clusters which are not 
       associated with tracks, including clusters from FD and SW but excluding the tagged electron.
       The same quality criteria are applied to all calorimeter clusters and charged tracks as in 
       the previous analysis\cite{ref:f2opal}.
       The masses of all particles in the hadronic system are assumed to be equal to the pion mass.
 \item $N_{\rm ch}> 2$, where $N_{\rm ch}$ is the number
         of charged particle tracks originating from the hadronic
         final state.
 \item The transverse momentum component \pbal\ of the event parallel to the tag plane has to be  lower than 3 \gev.
       The tag plane is defined by the momentum vectors of the incoming beam electron and
       the tagged electron.
       The transverse momentum of the event is the vector sum of the momenta perpendicular to the beam axis of 
       all measured particles, including the tagged electron.
 \item The transverse momentum component of the hadronic system perpendicular 
       to the tag plane
       \pout\ has to be lower than 3 \gev.
 \end{enumerate}
 These requirements select 7112 events,
 corresponding to an integrated luminosity of $70.8\pm 0.2$~\invpb, with \qzm$=2.8$~\gevsq.
 Using sets of independent triggers, the trigger efficiency was evaluated 
 to be (98$\pm$2)$\%$.\par 
 The dominant background comes from two-photon events with a lepton pair in the final state.
 These events are simulated with the Vermaseren Monte Carlo generator\cite{ref:Vermaseren}.
 Background from other sources such as other QED processes with four fermions in the final state or the process \znhad\ is found to be below $0.5\%$.
 The total estimated background contamination in the selected sample is $(2.5\pm 0.2)\%$.
 Deep inelastic electron-photon scattering is simulated with the generators  HERWIG 5.18d\cite{ref:Herwig}, PYTHIA 5.722\cite{ref:Pythia} and F2GEN\cite{ref:Jason}. F2GEN includes two final state models, ``pointlike'' and ``peripheral'', to describe the hard and soft limit of the process (see Ref.~\cite{ref:f2opal} for details). All Monte Carlo events are passed through a detailed detector simulations program\cite{ref:gopal} and the same reconstruction and analysis chain as the real data events.\par
\begin{figure}
 \begin{center}
\mbox{\epsfig{file=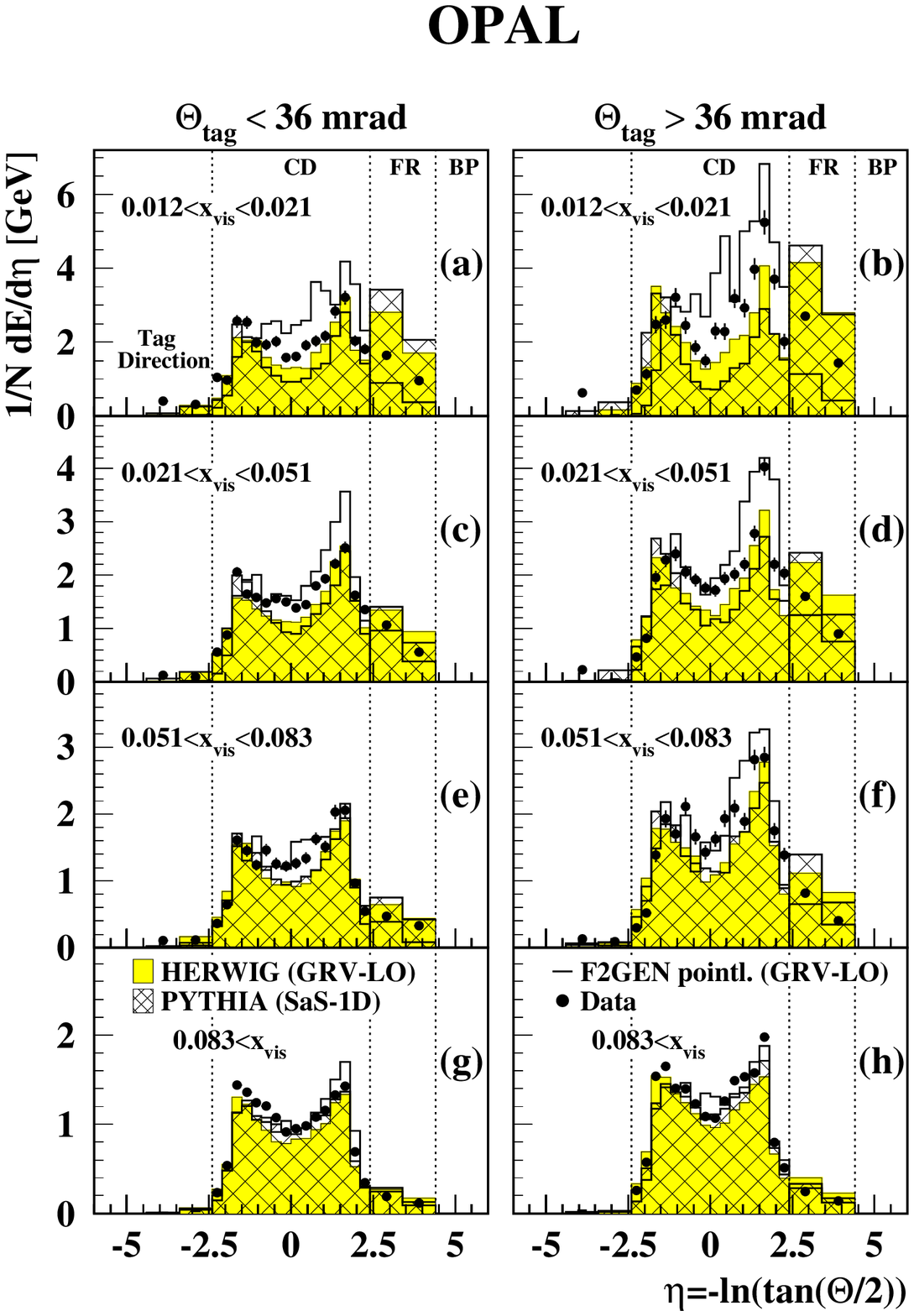,height=19 cm}}
 \caption{\label{fig:eta} 
 The hadronic energy flow per event as a function of pseudorapidity $\eta$ 
 for the data and various Monte Carlo samples. 
 In (a)-(f) the values of the bin limits in \xvis\ have been derived from the
 bin limits in $x_{\mbox{\tiny true}}$ of the three lowest $x$ bins in Figure \ref{fig:f2re}(a).  
 The bins in \ttag\ correspond approximately to the \qsq\ bins of 
 the unfolding.
 The errors shown are statistical only. 
 The vertical lines show the acceptance regions of the OPAL detector 
 components,
 CD = Central Detector, 
 FR = Forward Region and BP = Beam Pipe.
 (FR, BP not marked on the tag side).}
 \end{center}
 \end{figure}

 
 In Ref.~\cite{ref:f2opal} several distributions of measured final state hadronic quantities were studied in the \qsq~range $4.6\,\gevsq<\qsq<30\,\gevsq$, and significant differences were observed,
 both between data and the Monte Carlo models, and between the different Monte Carlo models.
 These distributions have been re-examined in the different kinematic range of this analysis and a similar
 pattern of disagreements is seen, especially in the energy flow distributions (Figure \ref{fig:eta}), and in the summed transverse energy perpendicular to the tag plane. These differences are found to be most prominent for $\xvis<0.05$, and small at higher \xvis.
 (The peripheral final state model of F2GEN has a similar behaviour to PYTHIA and is not shown.)
 Such uncertainties in the modelling of the hadronic final state give rise to uncertainties in the unfolded \ftxq~\cite{ref:f2opal}. They are included in the systematic error using the same set of Monte Carlo models as in Ref.~\cite{ref:f2opal}, except for the F2GEN models.
 For F2GEN, a weighted mixture of the two final state models available in the generator is used throughout the analysis. This mixture has been optimised to improve the final state description by a fit to the hadronic energy flow of the data in the lowest $x$ bin.
To account for the uncertainty of this `ad hoc' procedure a
variation of the fit by $\pm$ three times its error is included
in the evaluation of the systematic error.\par
%
%
%
%
 The program RUN by
 Blobel\cite{ref:Blobel} is used to unfold the structure function \ft\ from the measured \xvis\ distribution.
 To resolve very low $x$ values the unfolding is performed on a logarithmic 
 $x$ scale.\par
\begin{figure}
 \begin{center}
 \mbox{\epsfig{file=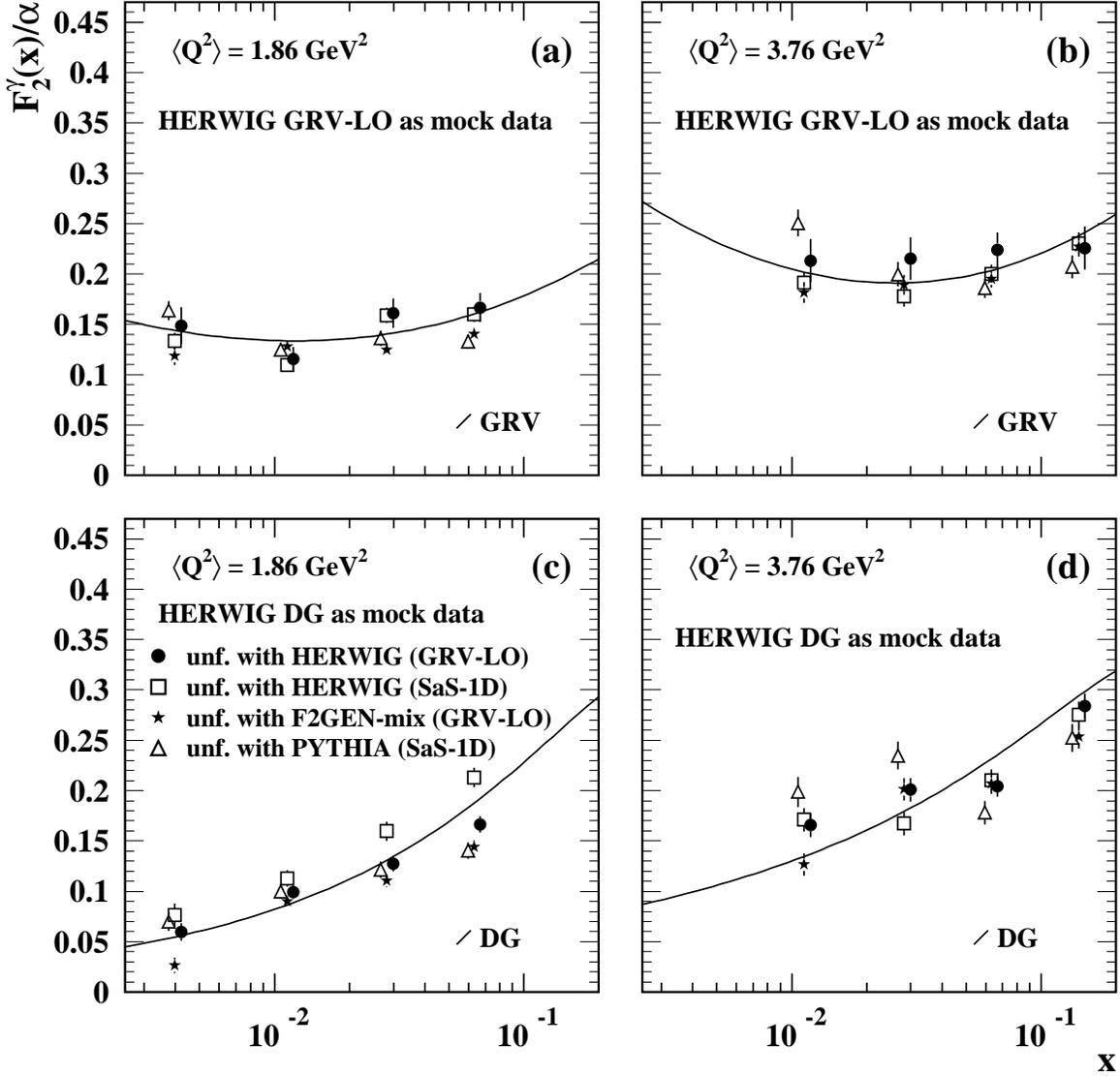,height=17cm}}
 \caption{\label{fig:chck} 
 Unfolding tests: In (a) and (b) mock data samples of HERWIG events with the GRV-LO parton density functions were unfolded using one of the four listed
 unfolding Monte Carlo samples. The solid curves show the GRV-LO \ftx\ for $\qsq=1.86\,\gevsq$ and $\qsq=3.76\,\gevsq$. In (c) and (d) a similar exercise was performed with mock data samples from HERWIG with the DG parton density functions.
 The solid curves show the corresponding DG values for \ftx\ at $\qsq=1.86\,\gevsq~$ and $\qsq=3.76\,\gevsq$.
 The error bars are statistical only. The symbols are slightly shifted in $x$ to avoid overlap.}
 \end{center}
 \end{figure}

 The ability of the unfolding to recover 
 the underlying structure function \ft\ of
 the data is tested by unfolding the known structure function of Monte Carlo 
 samples, the ``mock data'', instead of measured data.
 These mock data samples are then unfolded 
 with other Monte Carlo sample. 
 The unfolded \ft\ functions are compared to the original structure functions 
 used in the generation of the mock data sample.
 Figure~\ref{fig:chck}(a) and (b) show unfoldings of a mock data sample generated 
 with the HERWIG generator and the GRV-LO\cite{ref:GRV} parton density functions 
 using unfolding Monte Carlo samples from different generators and different input structure functions. The samples are
 divided into two \qsq\ bins with (a) \qsq~$<2.5$~\gevsq\ and (b)~\qsq~$>2.5$~\gevsq, respectively.
 The error bars include statistical errors only. The solid line is
 the GRV-LO structure function for the average \qsq\ of the mock data sample.
 Figure~\ref{fig:chck}(c) and (d) show similar plots for a mock data sample generated
 with HERWIG and the DG\cite{ref:DG}
 structure function. Here, the solid line is the DG structure function.\par
 The deviations observed represent the systematic impact on the unfolded result 
 of different parton density functions 
 and modelling of the hadronic final state.
 The DG structure function is used as a test for an
 \ft\ function that vanishes for $x \rightarrow 0$.
 The DG parton density functions have been evolved by their authors from $\qnq\, = 4.0$~\gevsq\cite{ref:DG}
 and are not supposed to be valid at lower \qsq, but they are suitable for this purely technical purpose. 
 Figure~\ref{fig:chck}(a) and (c)  demonstrate that despite the systematic errors 
 originating from different Monte Carlo models a structure function falling for $x \rightarrow 0$
 could be measured and distinguished from the GRV-LO prediction in the lower \qsq\ bin.
 In the higher \qsq\ bin the systematic errors are larger:
 The spread of the unfolding results shown in Figure~\ref{fig:chck}(d) does not allow the exclusion 
 of the \ft\ function described by GRV-LO.\par
%
%
%
%
%
%
%
 To determine the central values of the measured \ft\  
 a ``reference'' unfolding is defined.
 It is based on a HERWIG sample  
 generated using the 
 GRV-LO parametrisation, chosen
 for consistency with previous OPAL results\cite{ref:f2opal}.
 The event selection 
 as described above is applied.
 The data are divided into two \qsq~bins with \qsq~$<2.5$~\gevsq\
 and \qsq~$>2.5$~\gevsq\ which are unfolded separately. The two bins contain approximately equal numbers of events.
 The $x$ binning is chosen to keep correlations low between the unfolded \ft\ values in the different $x$ bins.\par
 The Monte Carlo generators HERWIG and F2GEN predict mean values of $\pzm = 0.03-0.08\,\gevsq$ for the virtuality of the quasi-real photon, depending on the model and the structure function used. 
 Several theoretical predictions exist for how \ft\ should behave 
 as a function of \psq~\cite{ref:GRVP2,ref:SASP2,ref:DGP2}.
 An estimate for the effect of the non-zero virtuality \psq\ yields an increase of \ft\ by roughly $10\%$,
 based on the \psq\ dependent structure function parameterisation of Schuler and Sj\"ostrand\cite{ref:SASP2} and the reference Monte Carlo sample.
 As the distribution of \psq\ in the data and the correct 
 theoretical prescription are not known, no correction is applied to the results.\par
\begin{figure}
 \begin{center}
 \mbox{\epsfig{file=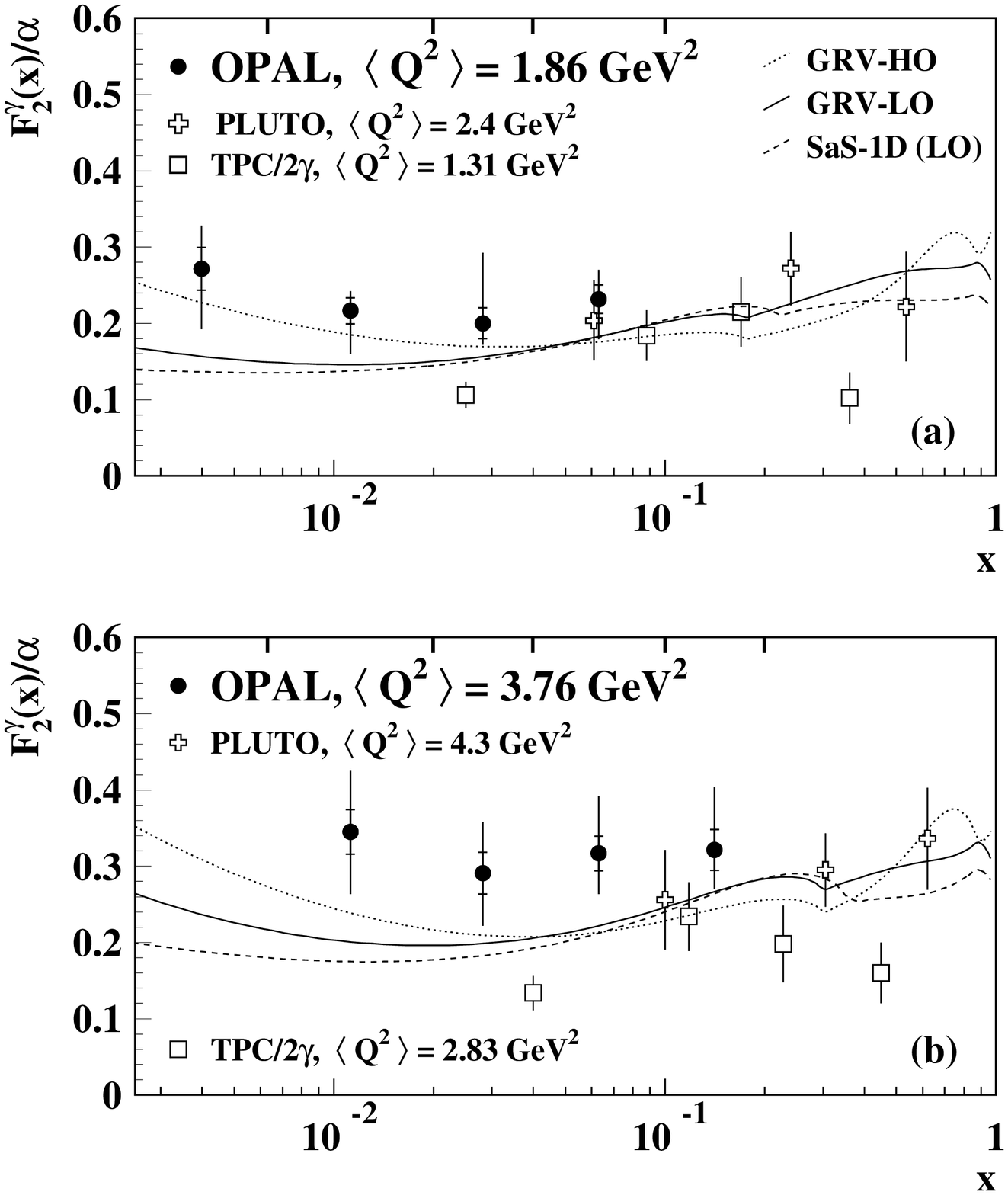,height=17cm}}
 \caption{\label{fig:f2re} 
 The full circles show our result for \ftx\ at $\qzm\,=1.86\,\gevsq$ (a) and $\qzm\,=3.76\,\gevsq$
  (b). 
 The total error and the statistical contribution are shown for each point.
 The tick marks at the top of the figure show the bin limits in $x$ for both \qsq\ ranges. These points are placed in the middle of the bin.
 The curves indicate the GRV-HO (dotted), GRV-LO (solid), and SaS-1D (dashed) predictions for \ft\ at the corresponding \qsq.
 The open symbols show results from other experiments at similar \qsq, with the total errors indicated only.
 These points are placed at the centre of the bins on a linear $x$~scale.
 }
 \end{center}
 \end{figure}

 \renewcommand{\arraystretch}{1.30}
 \begin{table}[htb]
 \begin{center}
 \begin{tabular}{|c|c|c|ll|c|c|}\hline
 \qsq\ & \qzm\ & \multicolumn{3}{c|}{$x$ range} & $x$ (centre of  & \ftn   \\
 \ [\gevsq] & \ [\gevsq]  & $-$log$_{10}(x)$ & \multicolumn{2}{c|}{$x$} & log$_{10}(x)$ bin) & \\\hline\hline
 $1.1-2.5$ & 1.86 & 2.6 -- 2.2 & 0.0025 & \hspace*{-0.4 cm} -- 0.0063 & 0.004  & \Z{0.27}{0.03}{0.05}{0.07} \\
             & & 2.2 -- 1.7  & 0.0063 & \hspace*{-0.4 cm} -- 0.020 & 0.011   & \Z{0.22}{0.02}{0.02}{0.05} \\
             & & 1.7 -- 1.4  & 0.020 & \hspace*{-0.4 cm} -- 0.040 & 0.028    & \Z{0.20}{0.02}{0.09}{0.02} \\
             & & 1.4 -- 1.0  & 0.040 & \hspace*{-0.4 cm} -- 0.100    & 0.063     & \Z{0.23}{0.02}{0.03}{0.05} \\\hline
 $2.5-6.6$   & 3.76 & 2.2 -- 1.7 & 0.0063 & \hspace*{-0.4 cm} -- 0.020 &  0.011 & \Z{0.35}{0.03}{0.08}{0.08} \\
             & & 1.7 -- 1.4 & 0.020 & \hspace*{-0.4 cm} -- 0.040 & 0.028 & \Z{0.29}{0.03}{0.06}{0.06} \\
             & & 1.4 -- 1.0  & 0.040 & \hspace*{-0.4 cm} -- 0.100 & 0.063     & \Z{0.32}{0.02}{0.07}{0.05} \\
             & & 1.0 -- 0.7 & 0.100 & \hspace*{-0.4 cm} -- 0.200 & 0.141 & \Z{0.32}{0.03}{0.08}{0.04} \\\hline
 \end{tabular}
 \caption{Results for \ft\ as a function of $x$
          in bins of \qsq. The first errors is statistical
          and the second systematic.
          The systematic errors do not contain the systematic effect caused by \psq\ being different from zero.}
 \label{tab:resxq}
 \end{center}\end{table}
 The unfolded \ftn\ for the data is listed in Table~\ref{tab:resxq}, and shown in Figure~\ref{fig:f2re}.
 The value of \ftn\ is given at the centre of each bin in log$_{10}(x)$. 
 The error bars show both the statistical error alone and
 the quadratic sum of statistical and systematic errors.
 Figure~\ref{fig:f2re} also shows the \ftn\ calculated from the GRV-LO
 and the SaS-1D\cite{ref:SaS1D}
 leading order parton 
 density parametrisations and the higher order GRV-HO\cite{ref:GRV} parametrisation, evaluated at the corresponding \qzm\ values.
 The parton density parametrisations given in Ref.\cite{ref:GS,ref:WHIT,ref:LAC,ref:FKP} are not shown because they are not supposed to be valid at these \qsq\ values, similarly to the DG parametrisation.
 In addition results from PLUTO\cite{ref:PLUTO} and TPC/2$\gamma$\cite{ref:TPC} for similar \qsq\ values are shown for comparison.\par
 The central values and statistical errors of the \ft\ measurements
 are determined using the reference unfolding. The statistical errors include, in addition to the statistical
 error of the measured data, also the error due to
 the limited number of Monte Carlo events (1.6 times the number of data events) 
 which is estimated to be $1/\sqrt{1.6}$ times the statistical error of the measured data.
 The estimation of the systematic errors includes the following
 four components:
 \begin{itemize}
 \item{The selection requirements are varied in order to change the signal and background event composition and to take 
 into account possible uncertainties in the simulation of variables which are used for the event selection.
 The size of the variations reflects the resolution of the measured
 variable and fulfils the requirement that the mean \qzm\ of the sample is
 not shifted significantly by the variation.
 }
 \item{An altered set of quality criteria for calorimeter clusters and tracks is used
 to determine systematic errors resulting from imperfections in the simulation 
 of the detector acceptance and calibration for tracks and calorimeter clusters.
 }
 \item{The unfolding is performed
 using the HERWIG generator and the standard selection, but replacing the GRV-LO parton density functions with SaS-1D
 in order to study the uncertainty due to the structure functions assumed in the 
 Monte Carlo samples.
 }
 \item{
 The unfolding is performed with the standard selection requirements using \linebreak
 PYTHIA and F2GEN
 in order to study 
 the effect due to a different modelling of the hadronic final state in the different Monte Carlo programs.
 } \par
\end{itemize}
 For each of the four systematic studies, the maximum deviations (above and below) 
 of the various unfolding results from the result of the reference unfolding 
 are taken as
 systematic errors. 
 The total systematic error assigned to the results in Table~\ref{tab:resxq}
 is the quadratic sum of these four contributions.\par
%
%
%
%
%
%
%
%
%
 In summary, OPAL data recorded in 1993 and 1994
 have been used
 to measure the photon structure function \ftxq\ at low $x$ using
 events with an electron tagged in the silicon 
 tungsten calorimeters.
 \ftxq\ has been unfolded 
 as a function of $x$ in two bins of \qsq, with $\qzm=1.86$ and $3.76\,\gevsq$.
 The unfolding has been performed on a logarithmic $x$ scale to resolve the
 lowest accessible $x$ values. 
 The data have been unfolded down to a minimum $x$ of 0.0025, 
 lower than measured previously\cite{ref:PLUTO,ref:TPC,ref:TASSO,ref:f2opal}. 
 This is the region where
 the proton structure function as measured at HERA starts to rise\cite{ref:HERA}.
 The extracted \ft\ result is consistent with the other OPAL \ft\ measurements\cite{ref:f2opal,ref:richard}.
 Within errors the results agree with the lowest value published by PLUTO\cite{ref:PLUTO}.
 Compared to the TPC/2$\gamma$ measurement\cite{ref:TPC} our results tend to be higher.
 Our result is consistent with a flat \ftx\ in both \qsq~ranges within the
 errors though it does not exclude a small rise with decreasing $x$.
The unfolded result is consistent in shape with the GRV-LO and SaS-1D parameterisations
for the corresponding \qsq\ values. However, the measured \ft\ is higher than
the GRV-LO and SaS-1D predictions.
 The GRV-HO prediction follows the data more closely and is in good agreement in the lower \qsq~bin.
%
%
%
%
%
%
%
%
%
%
%
%
\bigskip\bigskip

\noindent {\Large \bf Acknowledgements}
\par
\noindent We particularly wish to thank the SL Division for the efficient operation
of the LEP accelerator at all energies
 and for
their continuing close cooperation with
our experimental group.  We thank our colleagues from CEA, DAPNIA/SPP,
CE-Saclay for their efforts over the years on the time-of-flight and trigger
systems which we continue to use.  In addition to the support staff at our own
institutions we are pleased to acknowledge the  \\
Department of Energy, USA, \\
National Science Foundation, USA, \\
Particle Physics and Astronomy Research Council, UK, \\
Natural Sciences and Engineering Research Council, Canada, \\
Israel Science Foundation, administered by the Israel
Academy of Science and Humanities, \\
Minerva Gesellschaft, \\
Benoziyo Center for High Energy Physics,\\
Japanese Ministry of Education, Science and Culture (the
Monbusho) and a grant under the Monbusho International
Science Research Program,\\
German Israeli Bi-national Science Foundation (GIF), \\
Bundesministerium f\"ur Bildung, Wissenschaft,
Forschung und Technologie, Germany, \\
National Research Council of Canada, \\
Hungarian Foundation for Scientific Research, OTKA T-016660,
T023793 and OTKA F-023259.\\
%
 
%
%
%
%
%
%
%
 \pagebreak\bigskip
%
%
%
%
%
%
%
%
%
\end{document}